\documentclass[footinbib,twocolumn,amsmath,amstex,amssymb,mathfonts,superscriptaddress,prl,longbibliography]{revtex4-1}
\usepackage{amssymb}
\usepackage{mathrsfs}
\usepackage{graphicx}
\usepackage{float}
\usepackage[caption=false]{subfig}
\usepackage[pdftex,colorlinks=red]{hyperref}

\usepackage{epstopdf}

\usepackage[normalem]{ulem}
\usepackage{color}
\usepackage{bm}

\def\blue{\textcolor{black}}

\begin{document}

\def\qv{\vec{q}}
\def\magenta{\textcolor{magenta}}
\def\apricot{\textcolor{Apricot}}

\def\GJ{\textcolor{blue}}
\def\YH{\textcolor{Orange}}
\def\TT{\textcolor{ForestGreen}}

\newcommand{\norm}[1]{\left\lVert#1\right\rVert}
\newcommand{\ad}[1]{\text{ad}_{S_{#1}(t)}}

\title{Hybrid higher-order skin-topological modes in non-reciprocal systems}

\author{Ching Hua Lee} \email{calvin-lee@ihpc.a-star.edu.sg}
\affiliation{Institute of High Performance Computing, A*STAR, Singapore, 138632.}
\affiliation{Department of Physics, National University of Singapore, Singapore 117551, Republic of Singapore}
\author{Linhu Li}  \email{phylli@nus.edu.sg}
\affiliation{Department of Physics, National University of Singapore, Singapore 117551, Republic of Singapore}
\author{Jiangbin Gong}  \email{phygj@nus.edu.sg}
\affiliation{Department of Physics, National University of Singapore, Singapore 117551, Republic of Singapore}


\date{\today}
\begin{abstract}
Higher-order phases are characterized by corner or hinge modes that arise due to the interesting interplay of localization mechanisms along two or more dimensions. In this work, we introduce and construct a novel class of ``hybrid'' higher-order skin-topological boundary modes in  non-reciprocal systems with two or more open boundaries.  Their existence crucially relies on non-reciprocal pumping in addition to topological localization. Unlike usual non-Hermitian ``skin" modes, they can exist in lattices with \emph{vanishing} net reciprocity due to the selective nature of non-reciprocal pumping: While the bulk modes remain extended due to the cancellation of non-reciprocity within each unit cell, boundary modes experience a curious \emph{spontaneous breaking} of reciprocity in the presence of topological localization, thereby experiencing the non-Hermitian skin effect.
The number of possible hybridization channels increases rapidly with dimensionality, leading to a proliferation of distinct phases. In addition, skin modes or hybrid skin-topological modes can restore unitarity and are hence stable, allowing for experimental observations and manipulations in non-Hermitian photonic and electrical metamaterials.
\end{abstract}

\maketitle
Higher dimensions provide fertile settings for the fascinating interplay between qualitatively distinct phenomena. They not only support more sophisticated topological invariants, order parameters and dimensional reduction paradigms~\cite{ryu2006holographic,shi2006classical,schnyder2008classes,teo2010classes,ryu2010topological,RevModPhys.88.035005,gu2016holographic,qi2008}, but also allow lower-dimensional constituent phases to hybridize in novel ways~\cite{Zhang823,qi2008,xiao2017entangled}.
Of late, much focus has been on phases with higher-order (quadrupolar and beyond) topological polarizations
which, unlike conventional topological systems, support protected modes at the boundary of the boundary of a topological bulk~\cite{benalcazar2017quantized,benalcazar2017prb,langbehn2017HOTI,song2017HOTI,khalaf2018HOTI,kunst2018HOTI,ezawa2018HOTI,ezawa2018HOTI2,Schindlereaat0346,yichen2017HOTI,biye2018HOTI,li2018direct,Ezawaadd2018,Fulga2018}. Esoteric as they seem, such phases are already experimental realities in electronic, photonic, electrical and mechanical systems~\cite{erra-Garcia2018HOTI,christopher2018HOTI,imhof2018topolectrical,schindler2018HOTI,Ezawaadd2018}.

Inspired by recent experimental~\cite{lu2014photonics,tomoki2018photonics,malzard2015topo,malzard2018bulk,stefano2018pt,zhen2015spawning,el2018non,Zhang2018thermal,hofmann2018chiral} and theoretical~\cite{lee2016anomalous,alvarez2018non,xiong2018does,yao2018edge,kunst2018biorthogonal,yao2018non,kawabata2018non,lee2018anatomy,yin2018geometrical,jiang2018topological,lee2018tidal,liu2018second}
developments in non-Hermitian systems where boundary localization can also occur due to non-reciprocal pumping (the non-Hermitian ``skin'' effect \cite{yao2018edge,kunst2018biorthogonal,yao2018non}),
it is natural to ask if qualitatively new phases can emerge from the \emph{simultaneous} higher-order interplay between topological and non-reciprocal boundary pumping. As we shall elaborate, the answer is a firm affirmative: Not only do we find modes that are localized by both types of pumpings, we also uncover a novel mechanism of spontaneous breaking of {reciprocity} due to topological localization. As such, topological and non-topological modes are found to behave qualitatively differently in a higher-order setting, with the former still experiencing the skin effect even on a net-reciprocal lattice. 
These resultant modes have no Hermitian, reciprocal, or non-topological analog, and we shall henceforth refer to them as {hybrid} skin-topological (ST) modes. Without competing non-hybrid skin modes on a net-reciprocal lattice, these ST modes, which are also often real, are expected to dominate experimental measurements.





{\it Non-reciprocal boundary modes -- }
To convey the insights behind our hybrid ST mechanism, we first lay out a formalism describing the skin effect due to non-reciprocity.
Non-reciprocal lattices are characterized by unbalanced couplings $t_{a,b}\neq t_{b,a}$ between two lattice sites $a$ and $b$. When Hermiticity is also absent \footnote{Hermitian but non-reciprocal effects, i.e. magnetic fields, do not lead to boundary accumulation.} ($t_{a,b}\neq t_{b,a}^*$), the unbalanced couplings pump states towards their net direction en masse, leading to modified topological pumping~\cite{PhysRevB.23.5632,niu1990towards,PhysRevLett.77.570,soluyanov2011wannier,hu2017exceptional} as well as \emph{extensive} boundary accumulation known as the skin effect~\cite{xiong2018does,yao2018edge,lee2018anatomy}.
In contrast to topological boundary modes, the skin boundary modes are not topological, and
can emerge even in a one-band model.

\begin{figure}
\includegraphics[width=\linewidth]{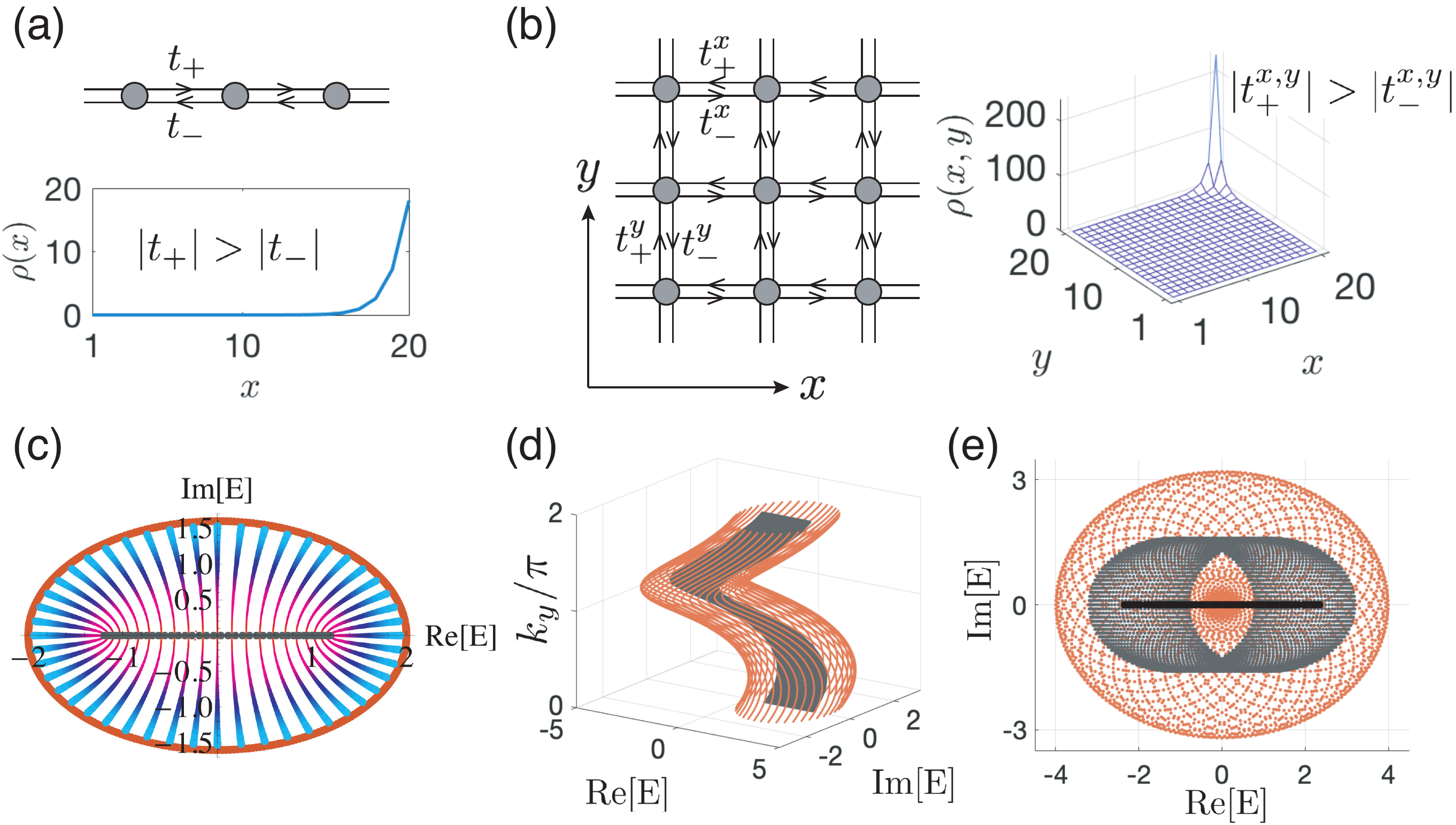}
\caption{(a) The 1D monoatomic non-reciprocal lattice of $H_\text{1D skin}$, and the squared skin eigenmode amplitudes summed over all eigenmodes \blue{$\rho(x)=\sum_{n}|\psi_n(x)|^2$} vs lattice site index $x$ with $t_{\pm}=1\pm0.8$.
(b) The 2D monoatomic non-reciprocal lattice of $H_\text{2D skin}$, and \blue{$\rho(x,y) =\sum_{n}|\psi_n(x,y)|^2$}  for $t^{x,y}_{\pm}=1\pm0.8$.
(c) PBC bulk spectrum (brown loop) of $H_\text{1D skin}$ with real OBC skin spectrum (gray) in its interior, blue-magenta curves representing the PBC-OBC (bulk-skin) interpolation.
(d) Spectra of $H_\text{2D skin}$ under double PBC (brown) and $x$-OBC/$y$-PBC skin modes (gray) in $E$-$k_y$ space.
(e) Double PBC spectrum (brown) and $x$-OBC/$y$-PBC (gray) spectra from (d), together with the real double OBC (black) spectrum, \blue{which is the result of the combined skin effects from both $x$ and $y$ directions.}
}
\label{fig:skin}
\end{figure}

The extent and direction of the skin effect experienced by an eigenmode under open boundary conditions (OBCs) can be quantified by the
\emph{magnitude and sign} of the decay length $L$, which can be obtained via complex analytical continuation of the Bloch momentum $k\rightarrow \tilde{k}=k+i\kappa$.
Specifically, for a momentum-space Hamiltonian $H(k)$, 
the inverse decay length $L^{-1}$ is given by the smallest $|\kappa|$~\cite{he2001exponential,lee2015free,lee2016band,lee2017band,xi2018hunting} for which
\begin{equation}
\text{Det}[H(k+i\kappa)-E\,\mathbb{I}]=0
\label{H2}
\end{equation}
is at least doubly degenerate in $\kappa$~\cite{lee2018anatomy}. Consider for illustration a 1D monoatomic chain along $\hat x$ with unbalanced couplings [Fig.~\ref{fig:skin}(a)]:
\begin{equation}
H_{\text{1D skin}}=\sum_x t_+\hat{c}_{x+1}^{\dagger}\hat{c}_{x}+t_-\hat{c}_{x}^{\dagger}\hat{c}_{x+1}
\label{1Dskin}
\end{equation}
with $H_{\text{1D skin}}(\tilde{k})=t_+e^{-i\tilde{k}}+t_-e^{i\tilde{k}}=E$. To solve for the skin modes $E$, we find
solutions $\tilde{k}_\mu,\tilde{k}_\nu$ of $H_{\text{1D skin}}(\tilde{k})=E$ with degenerate $\kappa_\mu=\kappa_\nu$. A short computation yields $e^{i\tilde{k}_{\mu,\nu}}=e^{ik_{\mu,\nu}}e^{-\kappa_{\mu,\nu}}=\left(E\pm\sqrt{E^2-4t_+t_-}\right)/(2t_-)$, from which $\kappa_\mu=\kappa_\nu$ yields the loci of the skin modes $E\in \mathbb{R}$, $|E|<2\sqrt{t_+t_-}$ [gray line in Fig.~\ref{fig:skin}(c)], which lies within the \blue{periodic boundary condition (PBC)} eigenenergy loci $E=t_+e^{-ik}+t_-e^{ik}$ [brown curve in Fig.~\ref{fig:skin}(c)].
In this simple model, all skin modes possess the decay length $L_{\text{1D skin}}=\left[\log\sqrt{|t_+/t_-|}\right]^{-1}$, which appear at the left/right boundary depending on whether $|t_-|>|t_+|$ or $|t_+|>|t_-|$. This is quantified by the summed squared eigenmode amplitude
\blue{$\rho({\bf r})=\sum_{n}|\psi_n({\bf r})|^2$}
(see Fig.~\ref{fig:skin}), where $\psi_n({\bf r})$ is the amplitude of the $n$-th eigenmode on site ${\bf r}$, and the summation runs over all eigenmodes.

Generically, OBC skin modes can be extrapolated from the PBC eigenmodes by mathematically increasing $\kappa$ from zero to $L_{\text{1D skin}}^{-1}$ i.e. along the blue-magenta curves of Fig.~\ref{fig:skin}(c). It can be shown that~\cite{lee2018anatomy} in the OBC limit, the (skin) spectrum [grey line in Fig.~\ref{fig:skin}(c)] converge along arcs or lines, which are exactly where $\kappa$ is doubly degenerate. Jumps in the biorthogonal polarization~\cite{kunst2018biorthogonal} occur when the skin arcs touch at $E=0$. The key take-home picture is that the skin effect (non-reciprocity) collapse the PBC spectra, which are generically closed loops, into open lines or arcs under OBCs. No $\kappa$ spectral flow and hence skin effect occurs in reciprocal systems, since their PBC spectra are already lines or arcs.

Fig.~\ref{fig:skin}(c) also suggests that the OBC skin spectrum can be real even when the PBC spectrum extends deep into the complex plane. The boundaries thus serve to restore unitary time evolution \blue{i.e. enforce probability-conserving time-evolution} by suppressing attenuation or instabilities along periodic spatial paths, which is very helpful experimentally. In this case, the reality of the skin spectrum follows from the observation that under OBCs,
the $H_{\text{1D skin}}(k)$ lattice is gauge equivalent to that of a Hermitian Hamiltonian $H_{\text{1D skin}}\left(k+i\log\sqrt{|t_-/t_+|}\right)$, which has  a real spectrum~\cite{yao2018edge,lee2018anatomy}. More generally, real skin spectra can be inferred from the symmetry of PBC loops under complex conjugation.


{\it Higher-order skin effect -- }
To complete the necessary formalism for describing hybrid ST modes, we next outline how to treat the non-reciprocal skin effect under multiple OBC directions. In such settings, non-reciprocal pumping leads to boundary skin mode accumulation along each OBC direction, leading to higher-order corner skin modes. 
For concreteness, consider a 2D generalization of our 1D non-reciprocal monoatomic lattice of Eq.~(\ref{1Dskin}) [Fig.~\ref{fig:skin}(b)]: 
\begin{equation}
H_{\text{2D skin}}(\bm k)=t^x_+e^{-ik_x}+t^x_-e^{ik_x}+t^y_+e^{-ik_y}+t^y_-e^{ik_y}.
\label{2Dskin}
\end{equation}
Under PBCs along both directions (double PBCs), the spectrum consists of a series of closed $k_x$ spectral loops parameterized by $k_y$, which collectively form a torus in the 3D space indexed by $(\text{Re}\,E,\text{Im}\,E,k_y)$, as shown in Fig.~\ref{fig:skin}(d).
OBCs in the $x$-direction ($x$-OBC/$y$-PBC) yield $k_y$-dependent 1D skin edge modes given by $E'=E-t^y_+e^{-ik_y}-t^y_-e^{ik_y}\in \mathbb{R}$, $|E'|<2\sqrt{t^x_+t^x_-}$, which form a strip-like spectrum within the double PBC torus of Fig.~\ref{fig:skin}(d). 
Introducing OBCs also in the $y$-direction (double OBCs), the 1D skin edge modes will also accumulate along the $y$-direction, forming 0D skin corner modes [Fig.~\ref{fig:skin}(b)] with spectra~\cite{SuppMat} $|E|<2(\sqrt{t^x_+t^x_-}+\sqrt{t^y_+t^y_-})$, $E\in \mathbb{R}$ [black line within gray loops in Fig.~\ref{fig:skin}(e)]. Analogous to 1D skin modes, 2D higher-order skin corner modes can be quantified by inverse decay lengths from the twice analytically-continued complex momentum, as  
elaborated in the Supplementary materials.

{\it Destructive interference of non-reciprocity -- }
\begin{figure*}
\begin{center}
\includegraphics[width= \linewidth]{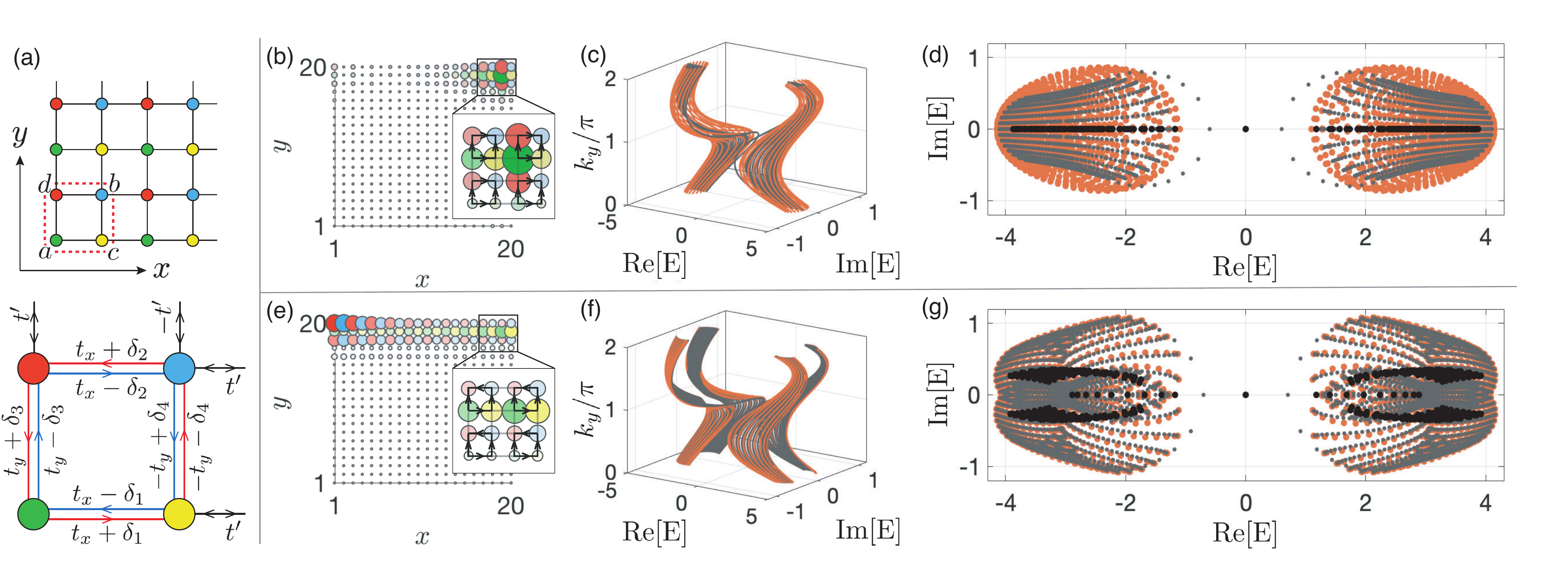}
\end{center}
\caption{
(a) Our 4-band model $\hat H_\text{4-band}$ with unit cell plaquette $(a,c,b,d)$, with $t_x=t_y=1$ and $t'=2$ as an example. (b-d) Corner (2nd-order) skin effect from constructive interference of non-reciprocities $\delta_{1,2,3,4}=0.4,-0.4,-0.8,0.8$, and (e-g) edge skin modes along $\hat y$ due to cancellation of $x$-direction non-reciprocities, with $\delta_{1,2,3,4}=0.4,0.4,-0.8,0.8$. (b,e) The eigenmode distribution, with darker and larger circles indicating larger summed squared amplitude $\rho(x,y)$ and sublattices indicated by color as in (a).
Black arrows in the insets represent the directions of net microscopic couplings.
(c,f) Double PBC (brown) and $x$-OBC/$y$-PBC (gray) spectra in $E$-$k_y$ space. (d,g) Spectra in (c,f) from above, together with double OBC (black) spectra, showing 0D skin corner modes (d) and 1D skin edge modes (g),  where the double PBC spectrum exhibit arcs instead of loops due to net $x$-direction reciprocity.
}
\label{fig:skin2}
\end{figure*}
To set the stage for ST hybrid modes, we minimally require a 2D lattice with 2 sites per unit cell in each direction. This allows sufficient degrees of freedom for topological localization as well as the key prerequisite destructive interference of non-reciprocity.
Consider a 4-band model with 2 sublattices in each direction $\hat x$ and $\hat y$: 
\begin{eqnarray}
\hat H_\text{4-band}=\sum_{\bm k}\hat\eta^\dagger_{\bm k} \left(\begin{matrix}
0 & 0 & H_{1,-} & H_{3,+} \\
0 & 0 & -H_{4,+}^* & H^*_{2,-} \\
H^*_{1,+} & -H_{4,-} & 0 & 0\\
H^*_{3,-} & H_{2,+} & 0 & 0
\end{matrix}\right)\hat \eta_{\bm k}\qquad
\label{model}
\end{eqnarray}
in the basis $\hat\eta^\dagger_{\bm k}=(\hat a^\dagger_{\bm k}, \hat b^\dagger_{\bm k},\hat c^\dagger_{\bm k},\hat d^\dagger_{\bm k})^T$, with $H_{j,\pm}=t_x\pm\delta_j+t' e^{-ik_x}$ for $j=1,2$ and $H_{j,\pm}=t_y\pm\delta_j+t' e^{-ik_y}$ for $j=3,4$. It is a mesh of 2 different non-reciprocal Su-Schrieffer-Heeger (SSH) models in each direction~\cite{lieu2018topological,yin2018geometrical,yao2018edge,su1980soliton} [Fig.~\ref{fig:skin2}(a)], and hence possesses the chiral symmetry $\bar\tau_z H_\text{4-band}(\bm k)\bar\tau_z=-H_\text{4-band}(\bm k)$, $\bar\tau_z=\tau_z\otimes\mathbb{I}$. 
The $\delta_j$'s quantify the non-reciprocal pumping influence in each acbd plaquette [Fig.~\ref{fig:skin2}(a)]; when they all vanish, $\hat H_\text{4-band}$ reduces to the well-studied Hermitian second-order topological model~\cite{benalcazar2017quantized,benalcazar2017prb,li2018direct}. The additional symmetry $H_\text{4-band}({\bm k})=H^*_\text{4-band}(-{\bm k})$ enforces $E({\bm k})=E^*(-{\bm k})$, leading to skin spectra symmetric about the real axis. Of them, many modes are real and exhibiting unitary time evolution, in analogy to those of the previously mentioned monoatomic lattice.


When non-reciprocities along both directions of each plaquette do not destructively interfere ($\delta_1\neq \delta_2,\delta_3\neq \delta_4$), we obtain skin corner modes [large blobs in Fig.~\ref{fig:skin2}(b)] like in our 1-band model. The double OBC spectrum (black) lies in the interior of the $x$-OBC/$y$-OBC spectrum (gray), which in turn lies in the interior of the double PBC spectral loops (brown), indicative of both $x$ and $y$-direction skin effects [Fig.~\ref{fig:skin2}(c,d)].
Due to nontrivial topological winding, a topological corner mode at $E=0$ is simultaneously present\cite{SuppMat}.

Consider next the case of Fig.~\ref{fig:skin2}(e), where the net non-reciprocities cancel along the $x$ but not $y$-direction ($\delta_1=\delta_2$, $\delta_3=-\delta_4$). We observe the skin effect only in the $y$-direction, as evident from the mode accumulation on the top but not the left or right edges. Additionally, topological corner modes are also present at the top left corner ($x=1,y=20$), although they do not scale extensively with system length, in contrast to the skin modes.  The $x$-direction skin effect is seen to be absent,
because the PBC spectra already form strips [Fig.~\ref{fig:skin2}(f)] or arcs [Fig.~\ref{fig:skin2}(g)] that preclude imaginary flux evolution and hence non-reciprocal pumping between the double-PBC (brown) and $x$-OBC/$y$-PBC (gray) spectra.

\begin{figure}
\begin{minipage}{\linewidth}
\includegraphics[width=\linewidth]{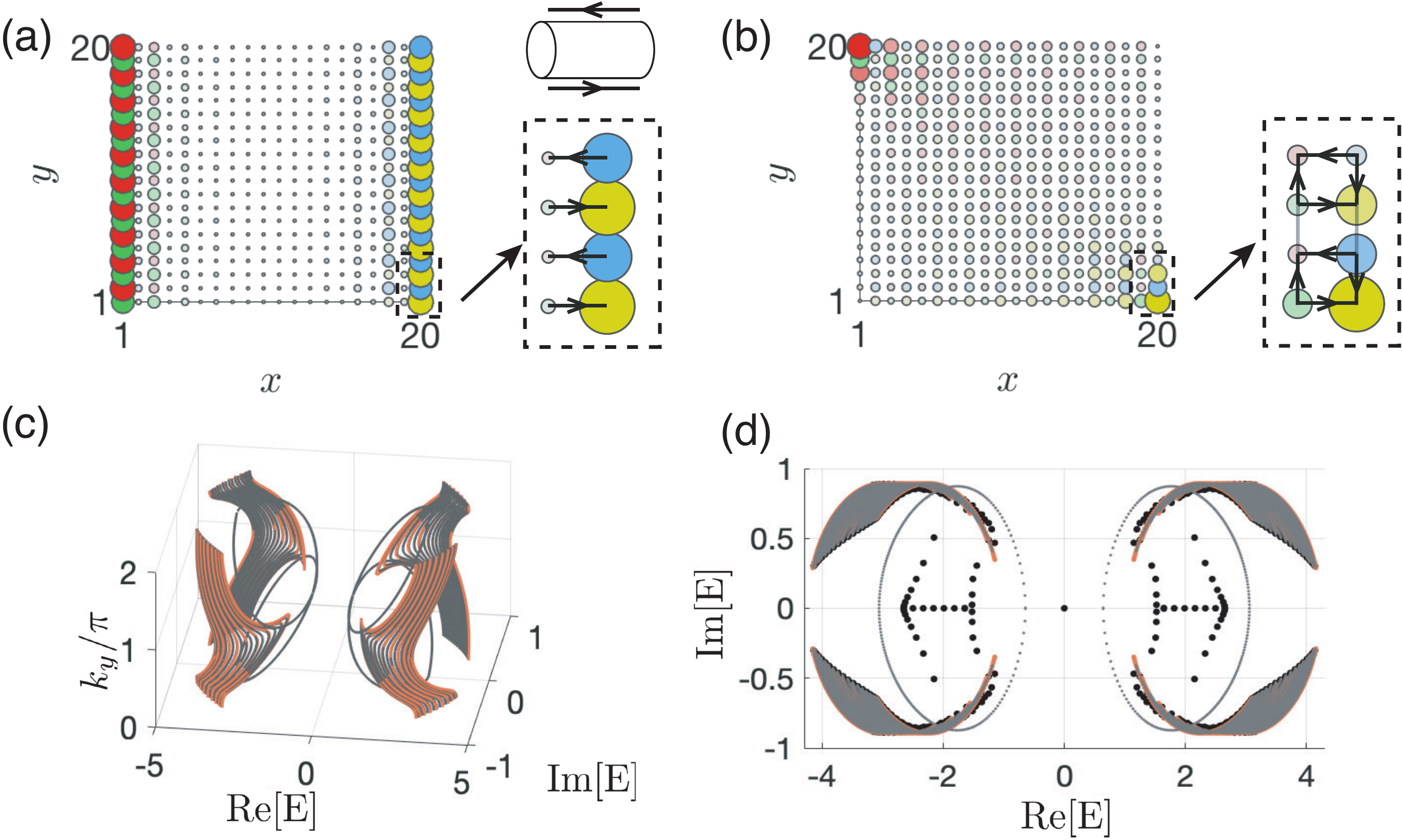}
\end{minipage}
\caption{(a,b) 
Spontaneous breaking of net reciprocity with topological modes, for $\delta_{1,2,3,4}=0.4,0.4,-0.8,-0.8$. (a) Density profile
of 1D topological edge modes $x$-OBC/$y$-PBC (cylinder geometry), indicating net \emph{local} non-reciprocity despite balanced non-reciprocity in each unit cell (black arrows represent net coupling non-reciprocity). (b) $\rho(x,y)$ plotted as in Fig.~2(b,e) shows hybrid ST modes under double OBCs. (c) Double PBC (brown) and $x$-OBC/$y$-PBC (gray) spectra in $E$-$k_y$ space, with distinct topological edge modes.
(d) The spectra of (c) together with double OBC (black) spectrum. Net reciprocity leads to no skin effect (arcs) for most modes, except for the topological mode loops  (gray) that produce hybrid ST modes (black) at their interiors~\cite{SuppMat}. }
\label{fig:skin2hybrid}
\end{figure}

{\it Hybrid skin-topological modes -- } We now describe the most intriguing and novel scenario of having hybrid skin-topological (ST) corner modes, whose existence requires \emph{both} topological protection and the skin effect. With couplings arranged such that the non-reciprocities cancel in both directions (Fig.~\ref{fig:skin2hybrid}), no skin effect is observed under OBCs in either direction, as expected from the vanishing net non-reciprocity. Yet, surprisingly, skin modes are still observed when OBCs are taken in both directions!

This conundrum is resolved by realizing that the double OBC corner modes [Fig.~\ref{fig:skin2hybrid}(b)] arise from the skin effect on the first-order topological modes [Fig.~\ref{fig:skin2hybrid}(a)] possessing unequal amplitudes on different sublattices. Not experiencing the full destructive interference of non-reciprocity, they are thus \emph{locally non-reciprocal} (Fig.~\ref{fig:skin2hybrid}b). Such spontaneous breaking of sublattice symmetry and hence non-reciprocity is generic among topological modes, and gives rise to a new class of \emph{hybrid} ST boundary modes.
Apparent in Fig.~\ref{fig:skin2hybrid}(c,d), both the double PBC (brown) and $x$-OBC/$y$-PBC bulk (gray) spectra consists of arcs (i.e. strips along $k_y$) and hence admit no skin effect. Yet, the $x$-OBC/$y$-PBC topological boundary mode (gray) is a locally non-reciprocal spectral loop which admits hybrid skin-topological modes arcs (black) in its interior~\cite{SuppMat}. As distinct from skin corner modes extensive in system area, hybrid ST modes scale with the system length. Furthermore, as one intriguing feature, hybrid ST modes emerge here with bulk modes being extended.  \blue{Though the skin effect itself is not topological, our analysis in the Supplementary Material~\cite{SuppMat} shows that
hybrid ST modes are featured by
the Chern number and a Berry phase \cite{SuppMat}.}

{\it Higher dimensional generalizations -- } In a generic $d$-dimensional lattice, each dimension can contribute skin (S) or topological (T) boundary modes, or neither (0). Hence there exist a total of $\binom{d+2}{2}-1=\frac{d(d+3)}{2}$ classes of nontrivial higher-order boundary modes, with the number of 0's giving the mode's spatial dimension. For instance, the $5$ classes of $d=2$ are T0, S0, TT, SS and ST, with the first two being topological and skin edge modes and the last three being corner modes. Of them, ST [Fig.~\ref{fig:skin2hybrid}] is the hybrid mode, while SS and S0 correspond to Figs.~\ref{fig:skin2}(b-d) and Figs.~\ref{fig:skin2}(e-g) respectively.

Of all these nontrivial classes, $\binom{d+2}{2}-1-2d=\frac{d(d-1)}{2}$ contain hybrid modes. For instance, the hybrid mode classes for $d=3$ are SST, STT and ST0, corresponding to skin-skin-topological, skin-topological-topological hybrid corner modes and skin-topological hybrid hinge modes respectively. Since the sequence for taking open boundaries is of no physical significance, permutations of T,S and 0 lead to no new classes. 
Being extensive, the total density of skin-topological modes scale with the system length $L$ like $L^s$, where $s$ is the number of S's.

\begin{figure}
\centering
\includegraphics[width=.9 \linewidth]{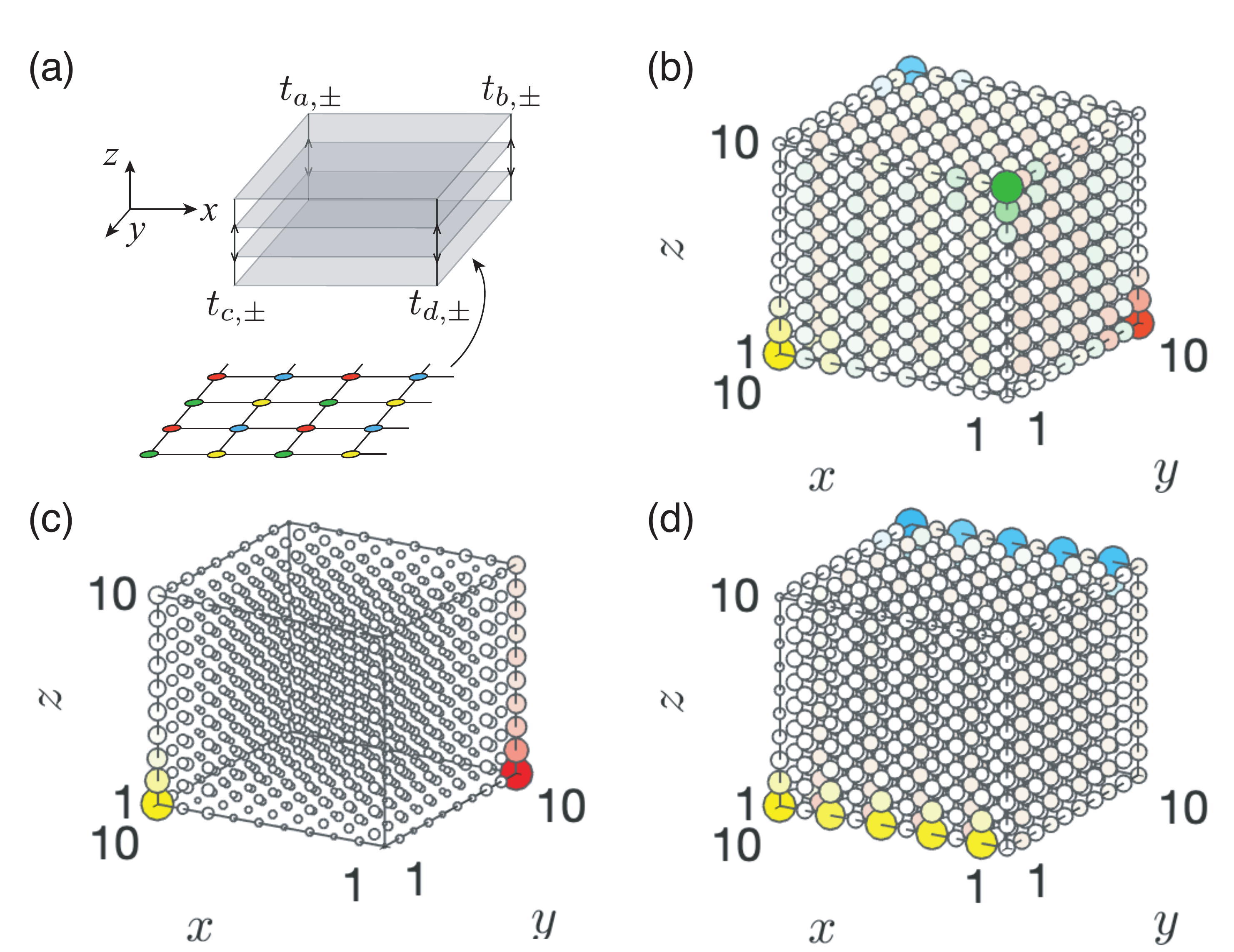}
\caption{(a) 3D lattice $\hat H_{3D}$ from stacks of 2D $\hat H_\text{4-band}$.  (b-d) Total site-resolved density $\rho(x,y,z)$ of corner and hinge modes, with darker and larger circles indicating larger normalized amplitudes, and color indicating sublattice localization. 
The parameters are $t_{x,y} = 1$ and $t'=2$ unless stated otherwise; and $\hat{z}$ couplings are given by $t_{\alpha=1}$, $\delta_{a,c}=-\delta_{b,d}=0.8$.
(b) STT corner modes from stacks of Hermitian 2D 2nd-order topological TT lattices, with $\delta_{1,2,3,4}=0$. (c) SST corner modes from stacks of non-Hermitian hybrid ST lattices, with $\delta_{1,2}=-\delta_{3,4}=0.8$.  {(d) ST0 hinge modes from stacks of Hermitian 2D weak topological (T0) lattices with 1D edge modes~\cite{li2018direct}, with $\delta_{1,2,3,4}=0$, and $t_y$ between $a,d$ sublattices is chosen to be $-5$.}
  }
\label{fig:3D}
\end{figure}

For concreteness, we realize the STT, SST and {ST0} classes with a 3D lattice comprising stacks of our 2D model $\hat H^z_\text{4-band}$ (Eq.~\ref{model}) at different heights $z$ [Fig.~\ref{fig:3D}(a)]:
\begin{eqnarray}
\hat H_{3D}=\sum_{z}\hat H^z_\text{4-band}+\sum_{x,y,z;\alpha,\pm}t_{\alpha,\pm}\hat{\alpha}^{\dagger}_{x,y,z}\hat{\alpha}_{x,y,z{\mp}1},\qquad
\end{eqnarray}
$t_{\alpha,\pm}=t_\alpha\pm\delta_\alpha$, $\alpha=a,b,c,d$ sublattices. Like before, we set the four $\delta_\alpha$'s to yield no net non-reciprocity along $\hat z$ ($\delta_{a,c}=-\delta_{b,d}$), so that resultant hybrid modes have no known analogs. Since $\hat H_{3D}$ is monoatomic in the $z$-direction for simplicity, it can only support the skin effect (S), and the other two possibilities of SS0 and SSS in this model are shown in the supplementary materials.

In Fig.~\ref{fig:3D}(b), we realize the STT phase by stacking 2D layers of Hermitian 2nd-order topological lattices (TT) with $\delta_{1,2,3,4}=0$. Its four topological corner modes are driven to the top and bottom layers, corresponding to the non-reciprocities of the hinges along $\hat z$. By stacking our hybrid 2D skin-topological systems (ST) with symmetrized non-reciprocities $\delta_{1,2}=-\delta_{3,4}=0.8$, we also obtain the SST phase [Fig.~\ref{fig:3D}(c)]. Its two groups of hybrid ST corner modes localize along a pair of diagonal hinges with both downward non-reciprocities, and are thus both driven to the bottom. Note the difference in normalized corner mode weightage between the STT and SST cases, whose mode densities scale like $L$ and $L^2$ respectively.
{Last but not least, the 2D model $\hat H^z_\text{4-band}$ with modified couplings has a weak topological insulating phase (T0)~\cite{li2018direct}, whose 1D edge modes combine to form 2D ($\hat x$-$\hat z$ or $\hat y$-$\hat z$) surface modes in the 3D system, and are pumped into 1D hinge modes by the skin effect along $z$ direction, as shown in Fig.~\ref{fig:3D}(d).}


{\it Discussion -- }
In non-reciprocal systems, boundary modes can arise from both topological and non-reciprocal pumping. By developing a formalism that treats both effects on equal footing, we investigated their exciting interplay in a higher-order setting. The highlight of this work is the discovery of hybrid skin-topological (ST) modes, whereby lower-order topological modes spontaneously break the net reciprocity and host higher-order hybrid skin-topological modes with no Hermitian or non-topological analog. A plethora of different hybridization classes exists in higher dimensions, epitomizing the exciting interplay of non-reciprocity, topology and boundary effects.
While reciprocity-breaking leads to complex spectrum in general, skin modes or hybrid skin-topological modes are of significant experimental interest because they can be real and are hence stable.
 \blue{It should be also fruitful to study disorder or interaction effects on the ST modes, with the expectation that disorder or interaction may
 also have an impact on the net reciprocity.}

\begin{acknowledgments}
{\sl Acknowledgements.--} C.H.L and L.L. contributed equally to this work.
J. G. acknowledges support by Singapore Ministry of Education Academic Research Fund Tier I (WBS No.~R-144-000-353-112) and by the Singapore NRF grant No.~NRF-NRFI2017-04 (WBS No.~R-144-000-378-281).
\end{acknowledgments}




%

\clearpage

\onecolumngrid
\begin{center}
\textbf{\large Supplemental Online Material for ``Hybrid higher-order skin-topological modes in non-reciprocal systems" } \\  \ \\
By Ching Hua Lee, Linhu Li and Jiangbin Gong
\end{center}

{\small This supplementary contains the following material arranged by sections:\\
\begin{enumerate}
\item Competition between non-reciprocal and topological localization through the SSH model
\item Detailed discussion on the second-order skin effect
\item Illustration of PBC-OBC interpolations, with special focus on the hybrid skin-topological modes
\item Topological characterization of hybrid skin-topological modes
\item {SS0 and SSS modes in the 3D model}
\end{enumerate}
}
\setcounter{equation}{0}
\setcounter{figure}{0}
\setcounter{table}{0}
\setcounter{page}{1}
\setcounter{section}{0}
\makeatletter
\renewcommand{\theequation}{S\arabic{equation}}
\renewcommand{\thefigure}{S\arabic{figure}}
\renewcommand{\thesection}{S\Roman{section}}
\renewcommand{\thepage}{S\arabic{page}}
\vspace{0.2cm}

\section{I. Non-reciprocal topological modes in 1D}

\begin{figure}[H]
\centering
\begin{minipage}{.65\linewidth}
\subfloat[]{\includegraphics[width=.34\linewidth]{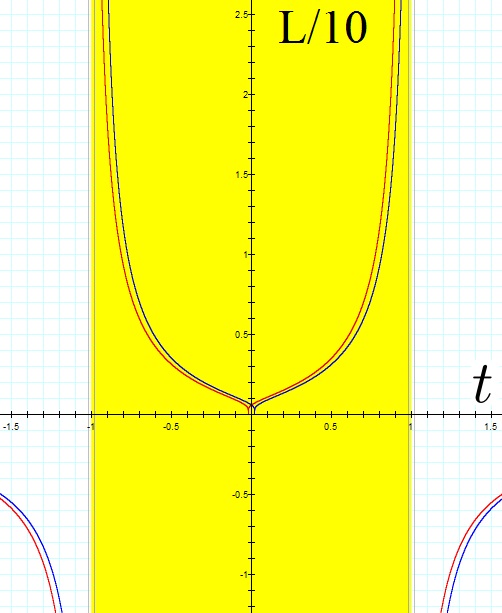}}
\subfloat[]{\includegraphics[width=.33\linewidth]{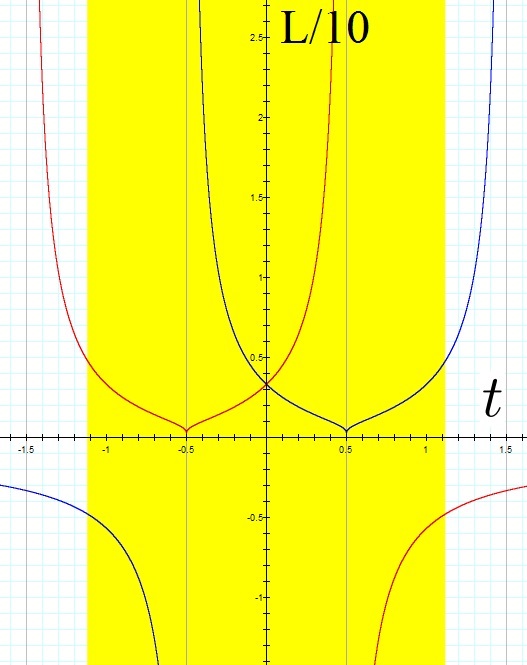}}
\subfloat[]{\includegraphics[width=.31\linewidth]{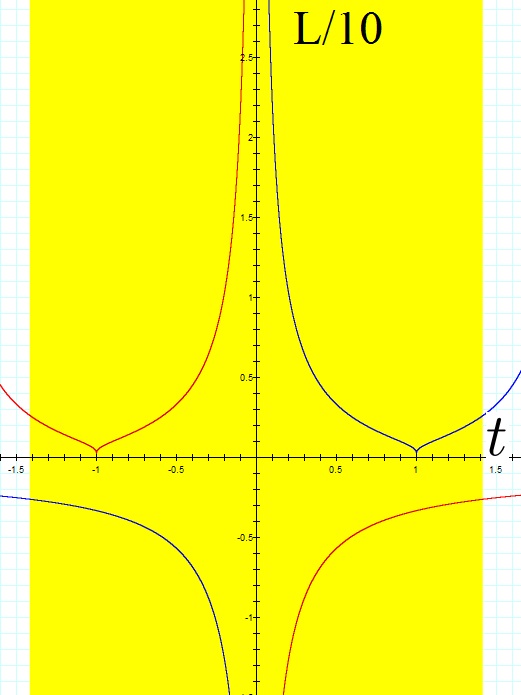}}
\end{minipage}
\caption{a-c) Plots of $L_{\text{left}}$ (blue) and $L_{\text{right}}$ (red) as a function of $t$ for $\delta=0.01,0.5,1$ and $t'=1$,  corresponding to a,b and c subfigures. The topological region is shaded in yellow. For very small $\delta$, the left and right modes have almost equal decay lengths. As $\delta$ grows, the decay length can become negative, which physically corresponds to a positive decay length on the other edge. At large $\delta$, i.e. $\delta=1$ (c), both edge modes exists at the left, leaving none on the right, even in the topological (yellow) region. }
\label{fig:corner1}
\end{figure}

In the main text, we have made references to topological modes in non-reciprocal systems, where a competition exists between topological localization and the skin effect. To illustrate this, and also to understand the 4-band model in the main text more deeply, consider the non-reciprocal Su-Schrieffer-Heeger (SSH) model
\begin{equation}
H_{\text{SSH}}(k)=(t+t'\cos k)\sigma_x+(i\delta+t'\sin k)\sigma_y
\end{equation}
with $\sigma_x,\sigma_y$ the Pauli matrices. The effect of non-reciprocity is to make the SSH chain ``look different'' from either end. Without $\delta$, a Hermitian SSH chain has a topological boundary mode at both ends if $|t|<|t'|$ (assuming sublattices with ABAB...AB termination), with both left/right boundary modes having a decay length $L=-\left[\log|\frac{t}{t'}|\right]^{-1}$. But when the non-reciprocal hopping $\delta$ is present, the non-Hermitian SSH chain becomes topologically nontrivial if~\cite{yin2018geometrical,yao2018edge,lee2018anatomy} $\sqrt{t^2-\delta^2}<|t'|$, with generically unequal left/right decay lengths
\begin{eqnarray}
L_{\text{left}}&=& -\left[\log\left|\frac{t-\delta}{t'}\right|\right]^{-1}\notag\\
L_{\text{right}}&=& -\left[\log\left|\frac{t+\delta}{t'}\right|\right]^{-1}
\end{eqnarray}
as shown in Fig.~\ref{fig:corner1}.
At small non-reciprocity $\delta$, we have $L^{-1}_{\text{left}/\text{right}}\approx \mp \log |t/t'|$, i.e. almost identically decaying modes on each boundary.
But at large $\delta$, both modes can accumulate on the \emph{same} boundary: the right/left boundary mode disappears and reappears on the left/right when $\pm t\,\delta>0$ and $|t\pm \delta|>|t'|$.



\section{II. Second-order skin modes}

Consider a 2D non-reciprocal Hamiltonian $H({\bm k})$ with ${\bm k}=(k_x, k_y)$,
1D edge skin modes are obtained by taking OBCs in one direction, say $\hat{x}$, and can be effectively represented by a complex non-Bloch momentum mode with inverse decay length $\kappa_{x}({\bm k})$ determined by Eq.~(2) in the main text. $\kappa_{x}({\bm k})$ generically depends on $\bm k$ because the PBC modes at different $\bm k$ may ``collapse'' into skin modes under different amounts of imaginary flux.
The effective Hamiltonian for $x$-OBCs is hence given by $H\left(k_x+i\kappa_x({\bm k}), k_y\right)$.
Likewise, if OBCs are also taken in the $y$-direction, resultant skin corner modes will be governed by the effective Hamiltonian
\begin{equation}
H(\tilde{\bm k})=H\left(k_x+i\kappa_x({\bm k}), k_y+i\kappa_y({\bm k})\right)
\label{H2nd}
\end{equation}
with the inverse decay length $\kappa_y({\bm k})$ similarly determined by Eq.~(2) in the main text. 
In higher dimensions, this procedure can be repeated ad infinitum until 0D corner modes with inverse decay lengths $\kappa_x({\bm k})$, $\kappa_y({\bm k})\,...$ are obtained.

\begin{figure}
\centering
\subfloat[]{\includegraphics[width=.6 \linewidth]{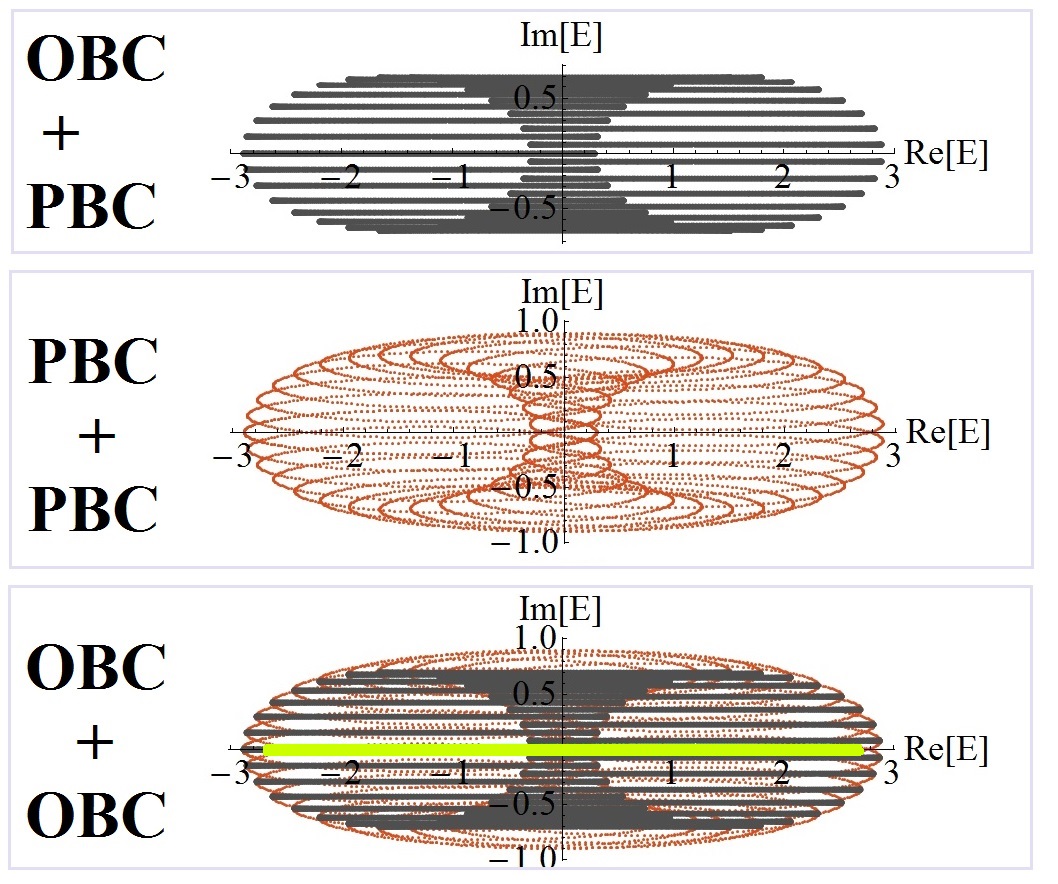}}
\caption{Higher-order spectrum of $H_{\text{2D skin}}$ under different boundary conditions. Parameters are $t^x_+=0.9,t^x_-=0.7,t^y_+=0.3,t^y_-=1$. In the bottom panel, the skin modes (yellow) are skin modes of the skin modes (gray) of the PBC modes (brown), and lie on the real axis due to reflection symmetry about it.}
\label{fig:skin2app}
\end{figure}

Here, we provide a more detailed derivation of the second-order skin mode results of the simplest illustrative model from Fig.~1(b) of the main text. It is a 2D monoatomic lattice model given by
\begin{equation}
H_{\text{2D skin}}(\bold k)=t^x_+e^{-ik_x}+t^x_-e^{ik_x}+t^y_+e^{-ik_y}+t^y_-e^{ik_y}.
\label{skin2Dapp}
\end{equation}
As explained, under PBCs along both directions (double PBCs), the spectrum consists of a series of closed $k_x$ spectral loops parametrized by $k_y$. Equivalently, it can also be considered as a series of closed $k_y$ spectral loops parametrized by $k_x$. As such, it traces out a projection of a torus in the complex $E$ plane. This is illustrated in the center panel of Fig.~\ref{fig:skin2app}, with model parameters slightly deformed from the main text for additional graphical clarity.

When taking boundary conditions, the system can be taken as a 1D model in the direction of the OBC, with the other momentum taken as parameters. From the main text on the 1D 1-band model, OBCs in the $x$-direction ($x$-OBC/$y$-PBC) yield $k_y$-dependent skin modes in 1D given by
\begin{align}
E'&=E-t^y_+e^{-ik_y}-t^y_-e^{ik_y}\in \mathbb{R},\notag\\
 |E'|&<2\sqrt{t^x_+t^x_-}.
\end{align}
As a parameter, $k_y$ modifies the effective energy $E'$ but does not affect $\kappa_x$. This can be seen in the top panel of Fig.~\ref{fig:skin2app}, where the $x$-OBC/$y$-PBC spectrum (gray) consists of straight lines (effective 1D skin modes) with centers displaced by $t^y_+e^{-ik_y}+t^y_-e^{ik_y}$. Further introducing OBCs also in the $y$-direction (double OBC), these effective 1D skin modes will accumulate at the top or bottom edges, forming the 0D skin corner modes. Geometrically, each eigenmode on the gray straight lines lies in an ellipse parametrized by $k_y$, and can thus still undergo another iteration of the skin effect pumping even though they already belong to an $x$-OBC spectra.

Repeating an almost exact computation as in the 1D derivation, but with $E$ replaced by $E'$, we obtain
\begin{align}
|E|&<2(\sqrt{t^x_+t^x_-}+\sqrt{t^y_+t^y_-}),\qquad E\in \mathbb{R}\end{align}
as the loci of the skin corner modes, which lies within in the interior of the $x$-OBC/$y$-PBC loops [yellow line within bottom panel of Fig.~\ref{fig:skin2app}]. Their spatial location depends on the sign of the decay lengths $L^v_{\text{2D skin}}=\kappa^{-1}_v=\left[\log\sqrt{|t^w_+/t^w_-|}\right]^{-1}$, $w=x,y$: e.g. skin modes accumulate on the top right corner if $L^x_{\text{2D skin}}>0$ and $L^y_{\text{2D skin}}>0$. 

\begin{figure}[H]
\centering
\subfloat[]{\includegraphics[width=.75 \linewidth]{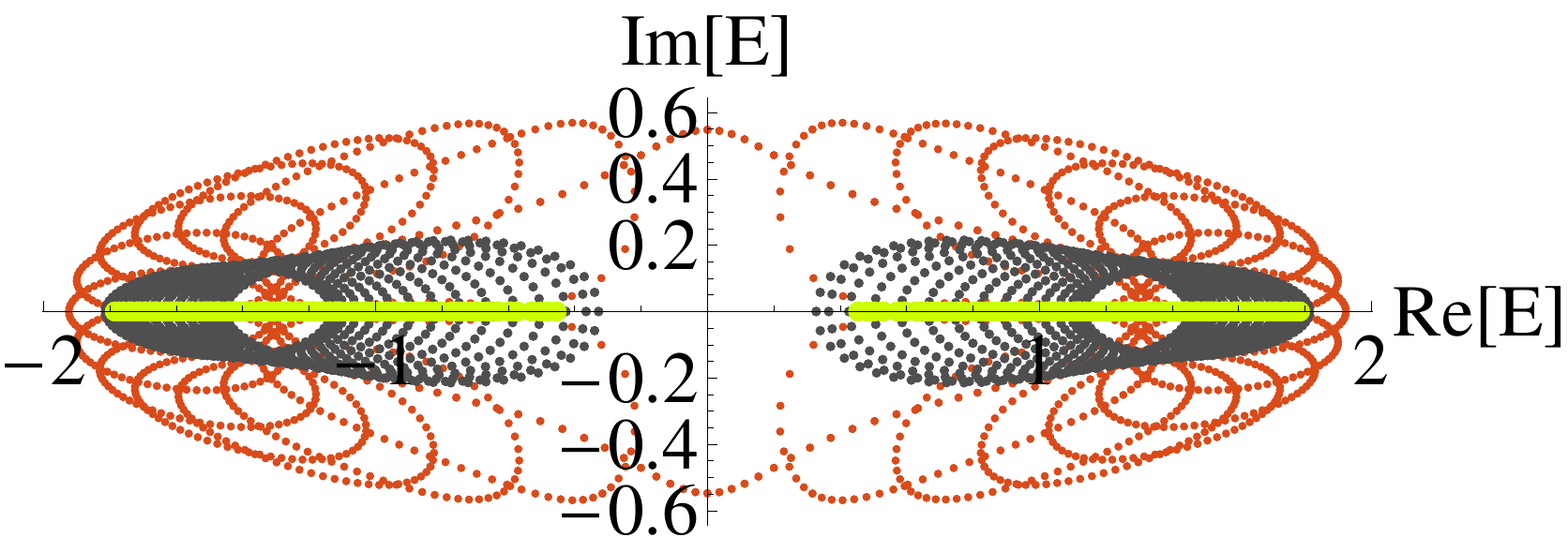}}
\caption{Double PBC (brown), $x$-OBC/$y$-PBC (gray) and double OBC (yellow) spectra of $H_\text{4-band}$ of the main text with $t'=0.5$, $t_x=t_y=1$ and $\delta_{1,2,3,4}=0.4,-0.4,-0.8,0.8$. The second-order skin effect is manifestly presented too, but with $\bold k$-dependent $\kappa_x(\bold k)$ and $\kappa_y(\bold k)$. }
\label{fig:2ndorderapp}
\end{figure}

Note that while the inverse decay lengths $\kappa_v$ are independent of Bloch momentum $\bold k$ in this simple model, in general they are not. Suppose we add further couplings beyond the nearest-neighbor terms $e^{\pm ik}$, or multiple bands, such that the PBC loops do not possess any geometric symmetry. Obviously then, their interior complex flux threading trajectories will no longer be symmetrical, and will terminate at degenerate arcs at different complex distances ($\kappa(\bold k)$) that are dependent on the original PBC starting point $\bold k$. An example with more than 1 band is shown in Fig.~\ref{fig:2ndorderapp}, which features the 4-band model $H_\text{4-band}$ from the main text with second-order skin corner modes, $t'$ adjusted to $0.5$ for greater graphical clarity. Its PBC loops cannot be easily deformed into rotationally-symmetric shapes, unlike ellipses, hence resulting in $\bold k$-dependent $\kappa_x(\bold k)$ and $\kappa_y(\bold k)$.

\section{III. PBC-OBC interpolations for $H_\text{4-band}$}

We now explicitly demonstrate how the PBC-OBC interpolation can be visualized using the technique of imaginary flux evolution, as first put forth by Ref.~\cite{lee2018anatomy}. To apply it, the key concept to understand is that the PBC-OBC evolution entails PBC eigenmodes moving from the PBC loop (with real momentum) into its interior, terminating only when the loop degenerates into arcs, lines or points. During this evolution, the eigenmodes necessarily become spatially localized (non-Bloch), because all extended Bloch states must lie on the PBC loop, where $\text{Im}\,\tilde k=\kappa=0$. Note that this evolution is only nontrivial in non-reciprocal lattices, since a reciprocal spectrum satisfies $E(k)=E(-k)$ and hence necessarily retraces back onto itself in an arc/line, which has no interior.
\begin{figure}[H]
\centering
\subfloat[]{\includegraphics[width=.6 \linewidth]{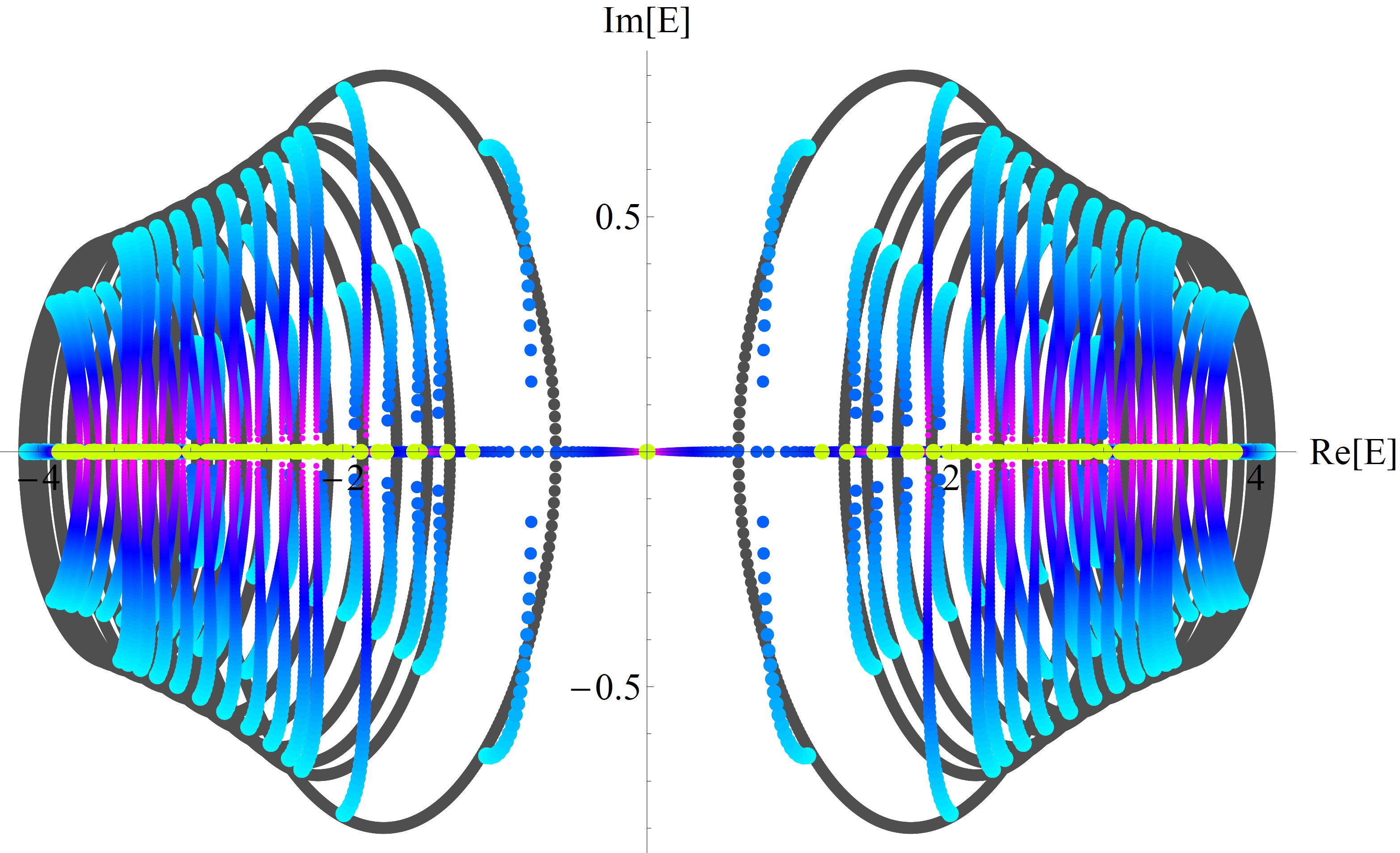}}\\
\subfloat[]{\includegraphics[width=.6 \linewidth]{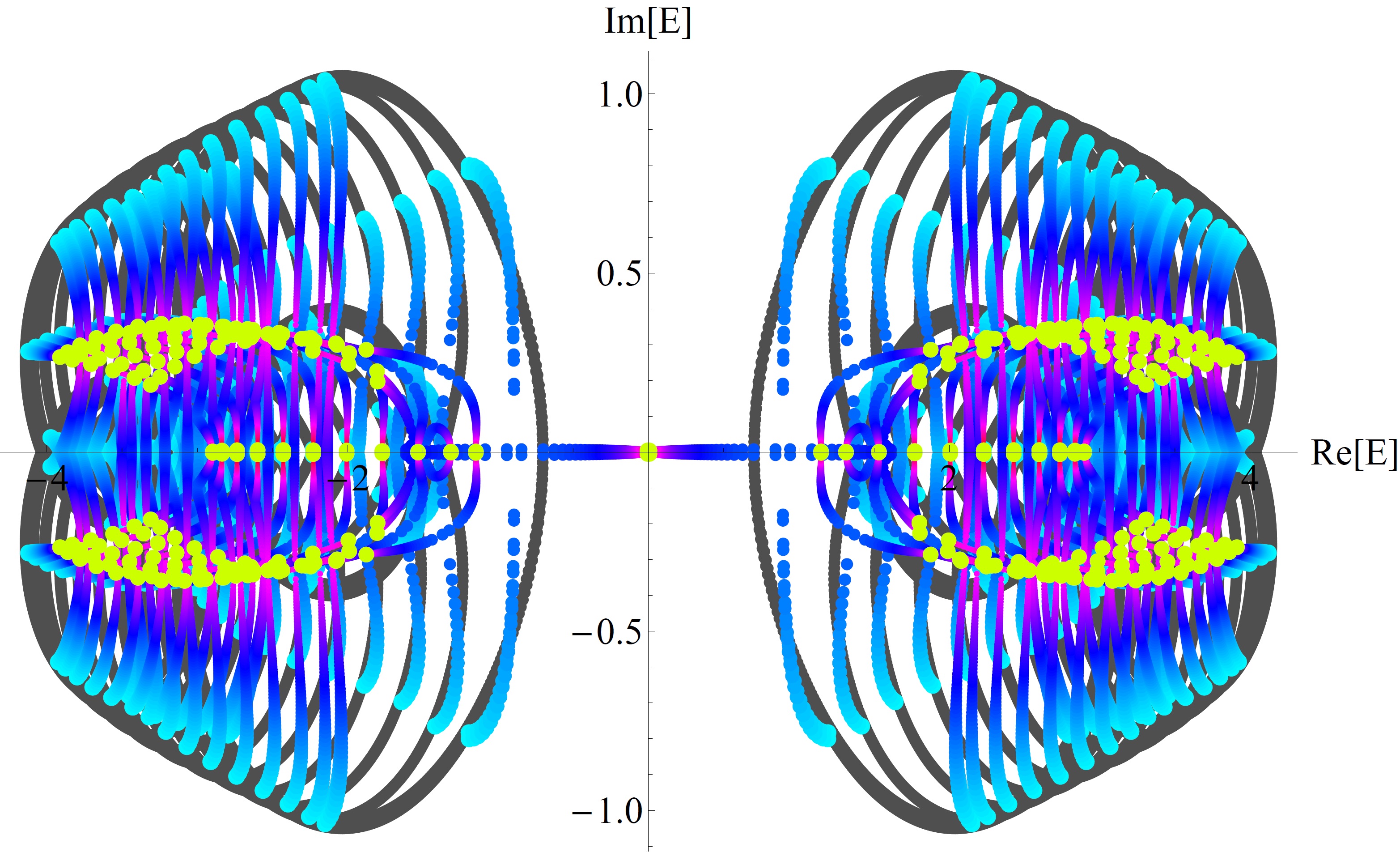}}\\
\subfloat[]{\includegraphics[width=.6 \linewidth]{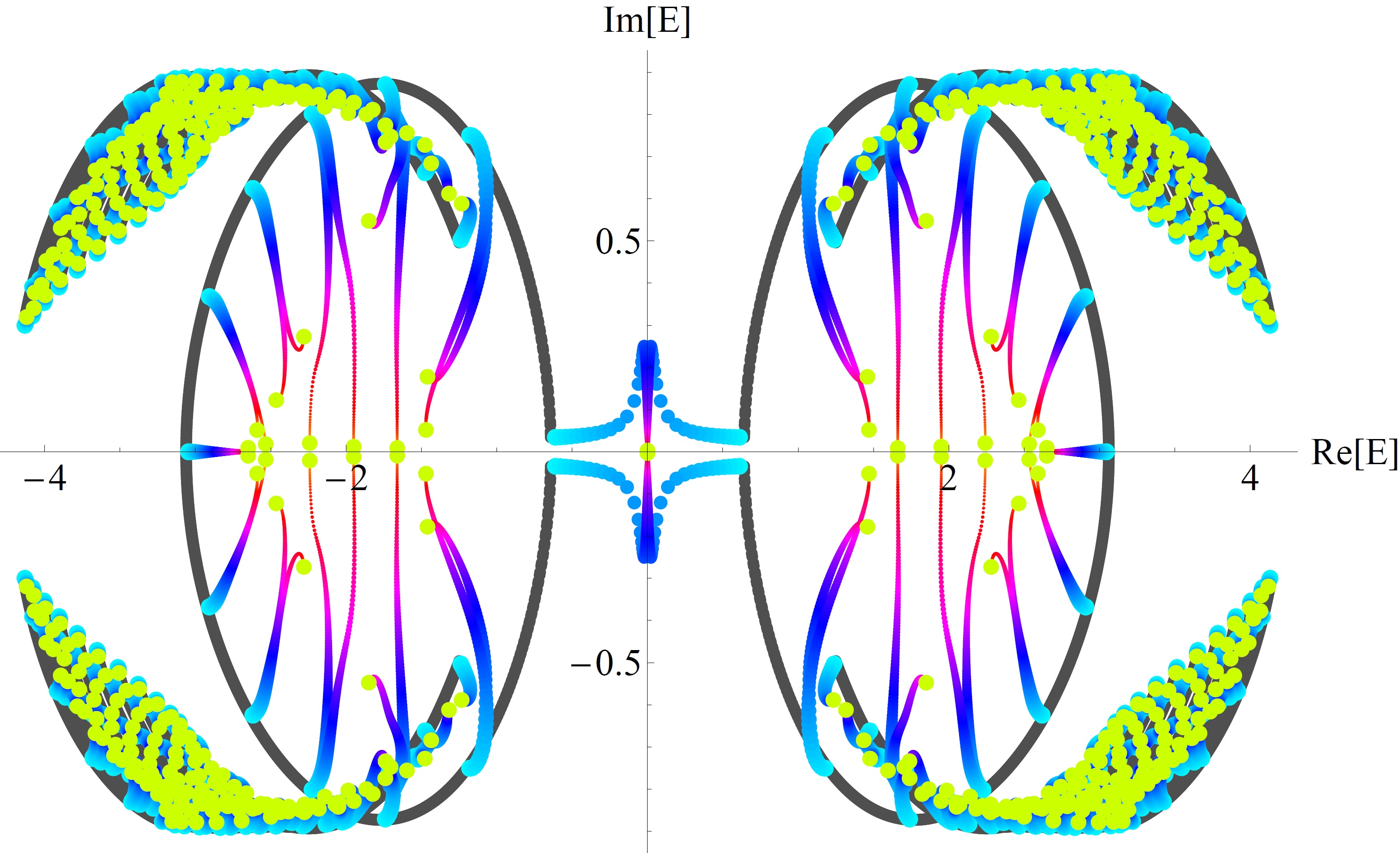}}
\caption{The $x$-OBC/$y$-PBC (gray) to double OBC (yellow) spectral flows (blue-purple) for the cases in Figs.~2d (a), 2g (b) and Fig.~3d (c) of the main text. Here the flow is taken with respect to the $y$-direction OBC, with the effective system being the 1D $y$-direction chain of $x$-OBC supercells. Notably, only the 1D topological edge modes (gray loops) exhibit imaginary spectra flow (skin effect) into the loop interior, leading to hybrid skin-topological modes.}
\label{fig:flows}
\end{figure}

\section{IV: Topological characterization of hybrid skin-topological modes}
\begin{figure}[H]
\centering
\includegraphics[width=0.9 \linewidth]{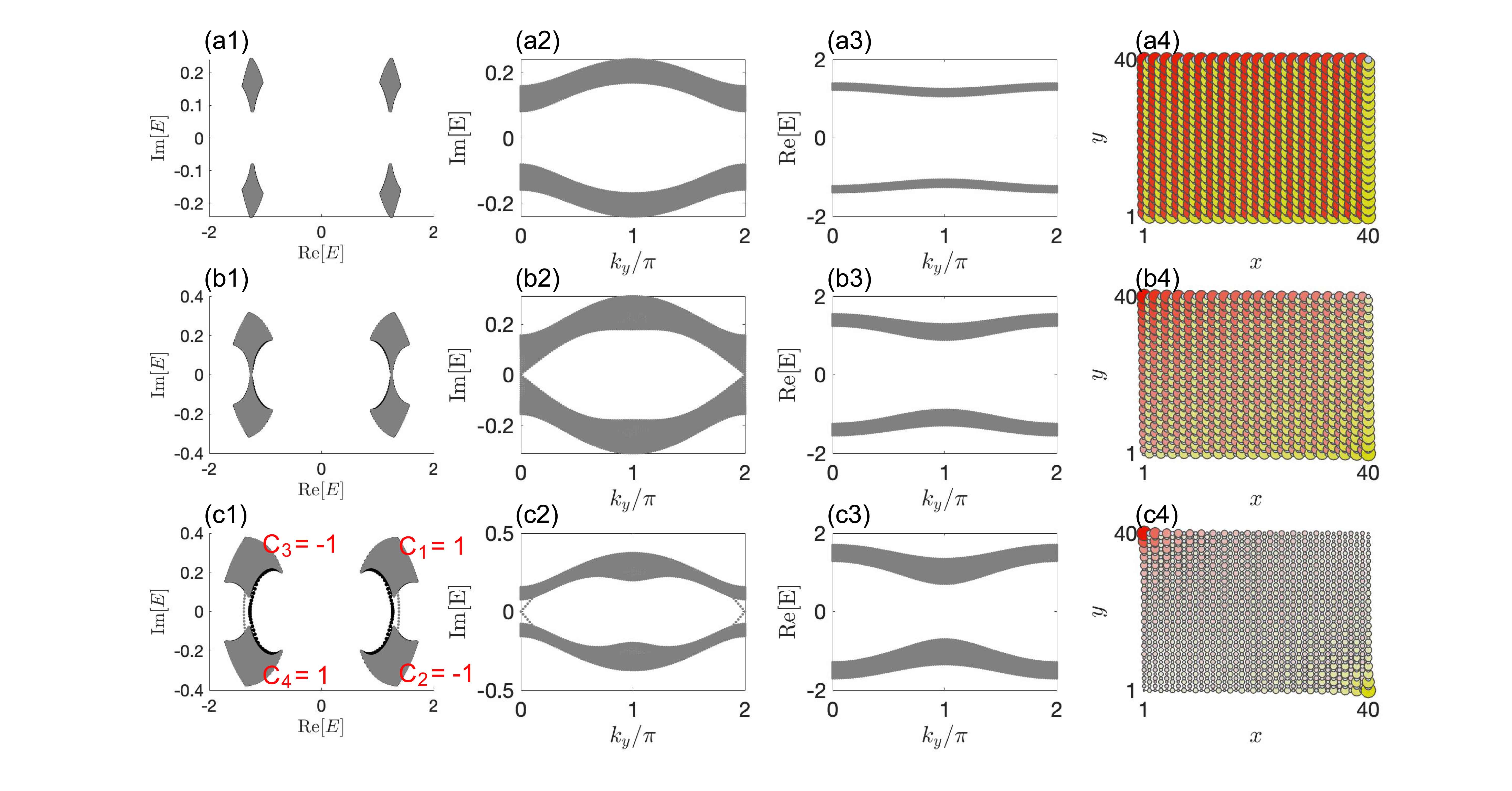}
\caption{The first column shows the x-OBC/y-PBC and double OBC spectra with gray and black colors respectively.
The second (third) column shows the imaginary (real) part of the x-OBC/y-PBC spectrum as a function of $k_y$.
The fourth column shows the summed squared eigenmode amplitude $\rho(x,y)$ under double OBC. The parameters are (a) $t'=0.1$, (b) $t'=0.2$, and (c) $t'=0.3$, with other parameters being $\delta_1=\delta_2=0.4$, $\delta_3=\delta_4=-0.6$, $t_x=t_y=1$. In (c1) we indicate the Chern number $C_n$ for each band.}
\label{fig:transition1}
\end{figure}

\begin{figure}[H]
\centering
\includegraphics[width=0.9 \linewidth]{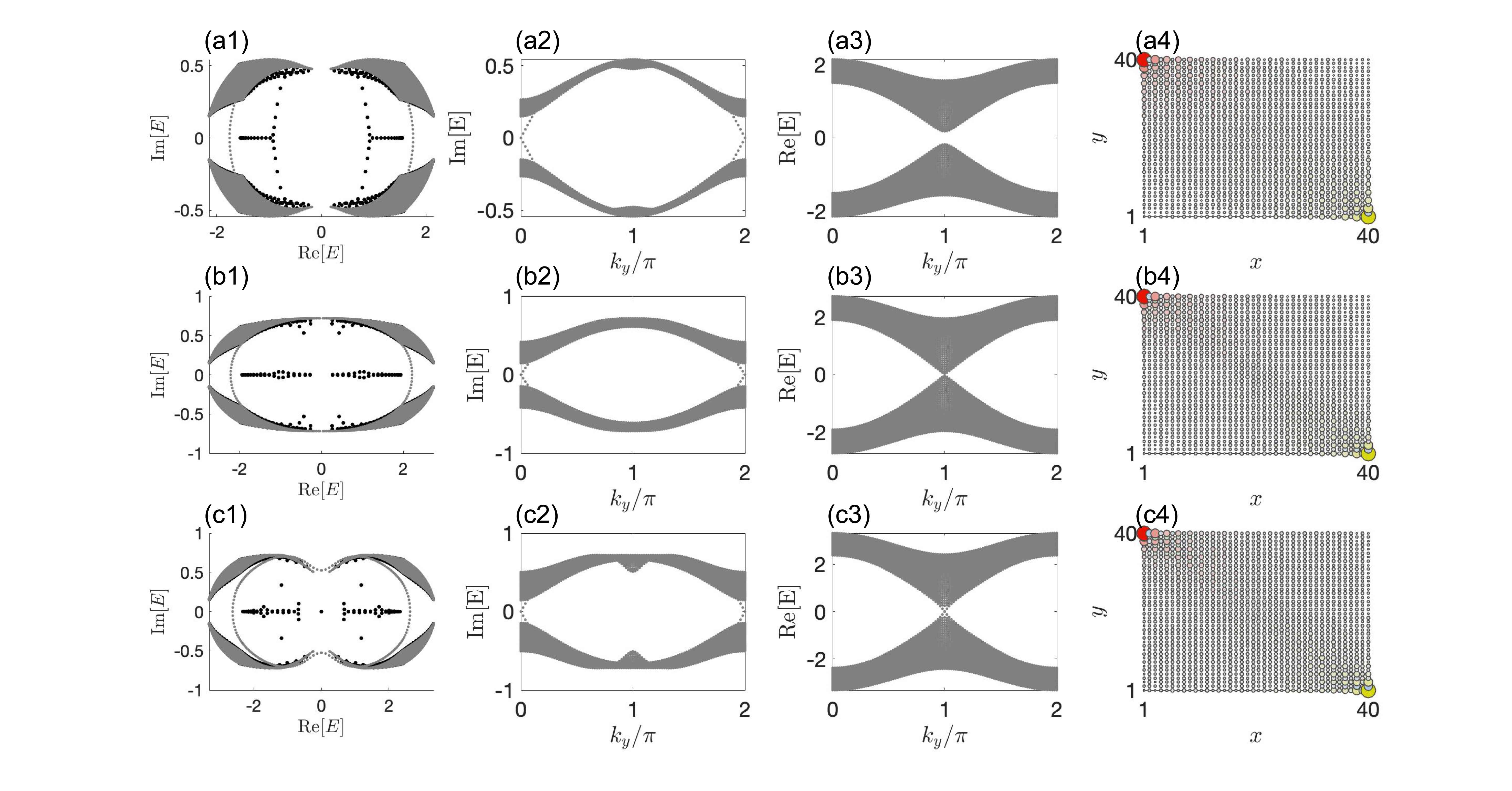}
\caption{The first column shows the x-OBC/y-PBC and double OBC spectra with gray and black colors respectively.
The second (third) column shows the imaginary (real) part of the x-OBC/y-PBC spectrum as a function of $k_y$.
The fourth column shows the summed squared eigenmode amplitude $\rho(x,y)$ under double OBC. The parameters are (a) $t'=0.6$, (b) $t'=1$, and (c) $t'=1.4$, with other parameters being $\delta_1=\delta_2=0.4$, $\delta_3=\delta_4=-0.6$, $t_x=t_y=1$.}
\label{fig:transition2}
\end{figure}

The existence of hybrid skin-topological modes requires both topological effect and non-Hermitian skin effect along the boundaries. In the case of Fig. 3 in the main text, the topological aspect of the hybrid corner modes arises from the 1D topological boundary modes under x-OBC/y-PBC. With some further analysis, we find that there exist at least two types of transitions as shown in Fig. \ref{fig:transition1} and \ref{fig:transition2}, where we fix the intracell hoppings and the strength of non-reciprocity, and change the value of the intercell hopping $t'$.

When $t'$ is small, the four bands are separated in the complex plane with no boundary mode connecting them, as shown in Fig. \ref{fig:transition1}(a). Increasing $t'$ will induce a topological phase transition at $t'\simeq 0.2$ for the parameters we choose, where the energy-bands touch each other at the real axis when $k_y=0$. After this transition, the system develops a pair of chiral-like 1D boundary modes in its imaginary spectrum [Fig. \ref{fig:transition1}(c2)], which then become the hybrid skin-topological corner modes shown in Fig. \ref{fig:transition1}(c4).
Defining the Chern number $C_n$ for the $n$th band as
\begin{eqnarray}
C_n=\frac{1}{2\pi}\iint d k_x dk_y (\partial_{k_x}A_{n,k_y}-\partial_{k_y}A_{n,k_x}),~A_{n,\alpha}=\frac{i\langle \psi_n^{L} |\partial_\alpha| \psi_n^{R} \rangle}{\langle \psi_n^{L} | \psi_n^{R} \rangle},
\end{eqnarray}
we find that $C_n$ takes the value of $\pm 1$ in the presence of the above boundary modes, as indicated on the figure panel.
On the other hand, we have $C_n=0$ for each band when topological 1D boundary states are absent and hence hybrid skin-topological modes are also absent. This is illustrated in Fig. \ref{fig:transition1}(a4).

\begin{figure}[H]
\centering
\includegraphics[width=0.8 \linewidth]{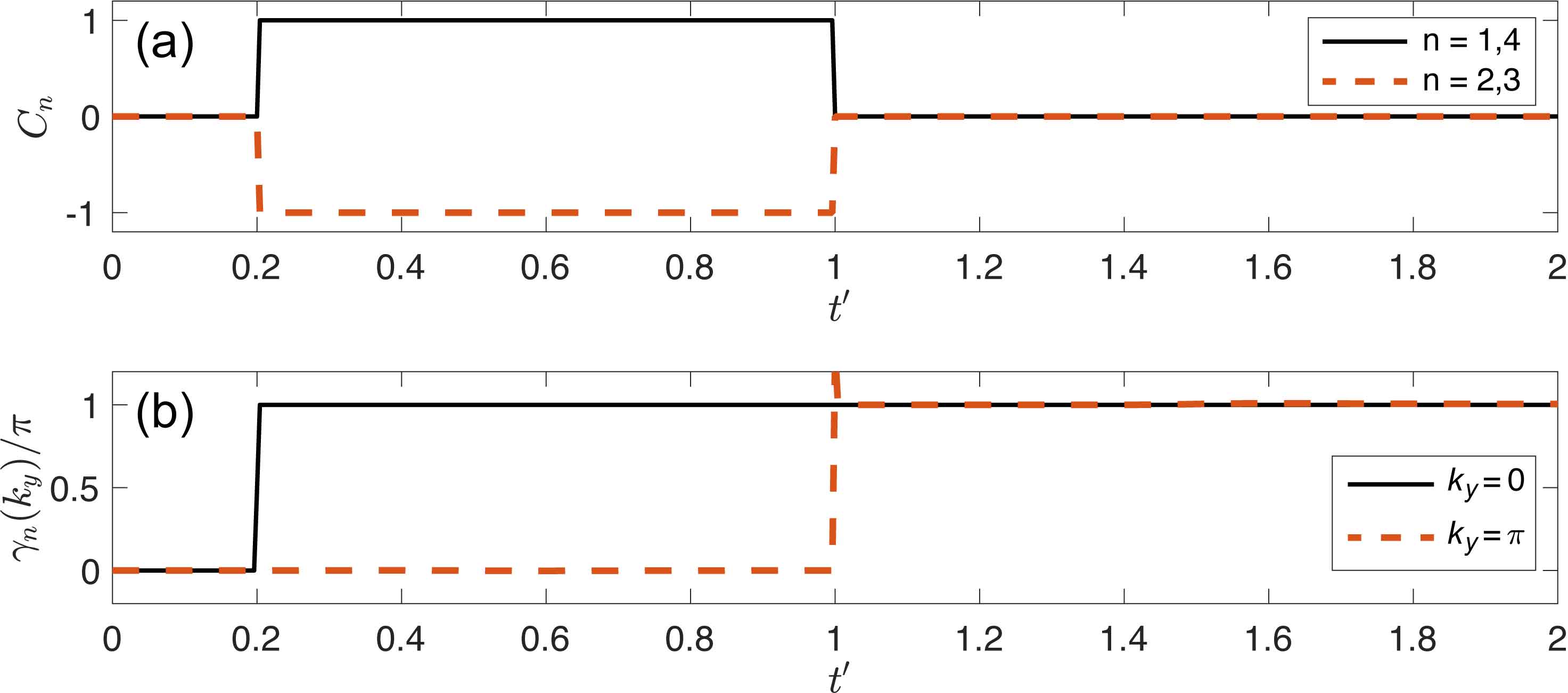}
\caption{The Chern number $C_n$ and Berry phase $\gamma_n(k_y)$ at $k_y=0$ and $\pi$, as functions of $t'$. In our model the Berry phase always takes the same value for different band index $n$. Other parameters are $\delta_1=\delta_2=0.4$, $\delta_3=\delta_4=-0.6$, $t_x=t_y=1$.}
\label{fig:phase_diagram}
\end{figure}

The second topological phase transition occurs at $t'\simeq 1$, where the energy bands touch each other at imaginary axis when $k_y=\pi$, as shown in Fig. \ref{fig:transition2}. After this transition, a pair of chiral-like 1D boundary modes emerges in its real spectrum [Fig. \ref{fig:transition2}(c3)]. The Chern number is found to be $C_n=0$ for each band in this case, suggesting that the two pairs of boundary modes correspond to opposite chiralities, and are no longer protected by a Chern topology.
Nevertheless, in this model, the above-mentioned  topological phase transitions are seen to occur only at $k_y=0,\pi$, where the possible boundary modes cross each other at zero (imaginary or real) energy. Such crossing boundary modes can be characterized by a Berry phase defined for an effective 1D Hamiltonian with $k_y$ taken as a parameter,
\begin{eqnarray}
\gamma_n(k_y)=i\int dk \frac{\langle \psi_n^{L} |\partial_{k_x}| \psi_n^{R} \rangle}{\langle \psi_n^{L} | \psi_n^{R} \rangle},
\end{eqnarray}
at $k_y=0$ and $\pi$. A $\pi$ Berry phase at either of these two points ensure the existence of 1D boundary states, and hence hybrid skin-topological modes under double OBC.
In Fig. \ref{fig:phase_diagram} we illustrate the phase diagram of our model regarding $C_n$, $\gamma_n(0)$, and $\gamma_n(\pi)$, which clearly indicates the above two topological phase transitions. Note that the second transition does not eliminate the 1D boundary modes under x-OBC/y-PBC, therefore it does not affect the hybrid skin-topological modes qualitatively.

\section{V: SS0 and SSS modes of the 3D model}

\begin{figure}[H]
\centering
\includegraphics[width=.6 \linewidth]{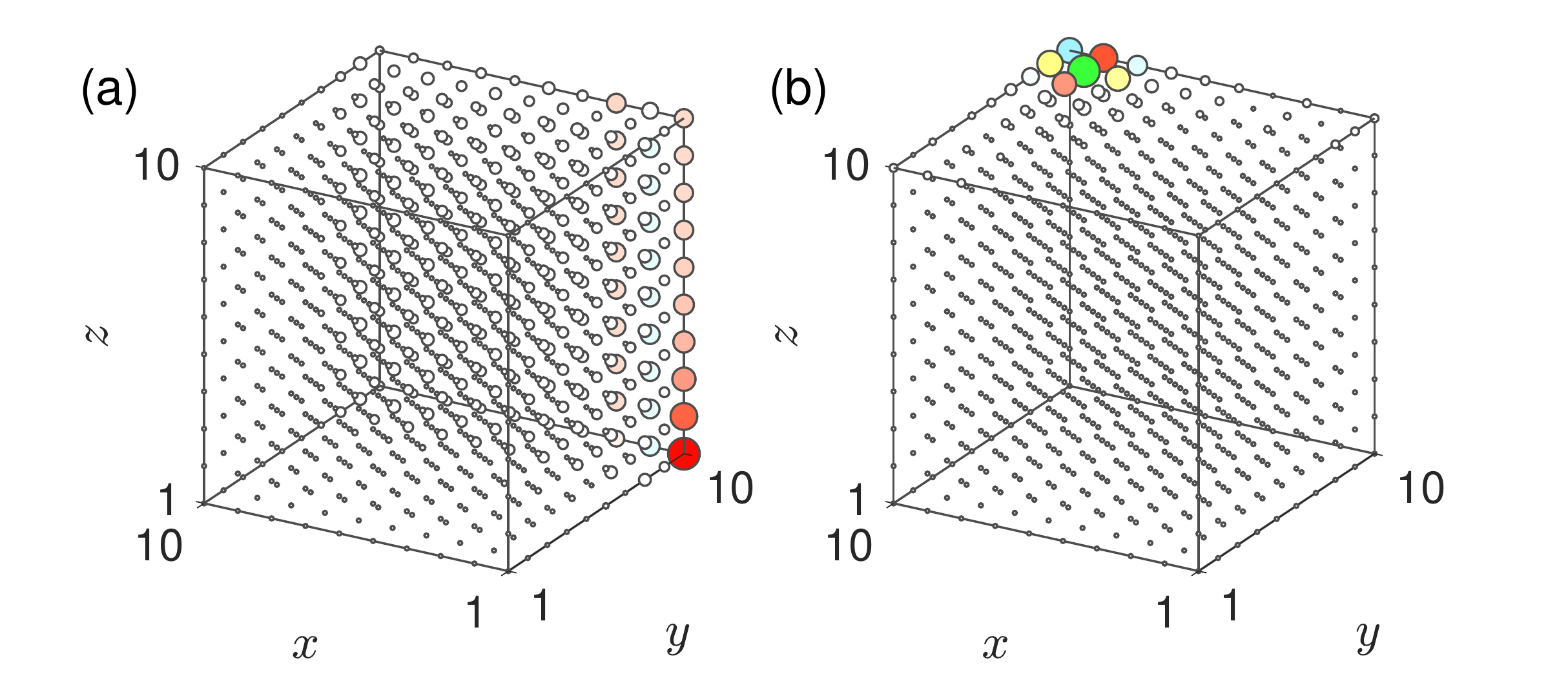}
\caption{The total site-resolved density $\rho(x,y,z)$ of 3D skin modes with $t_{x,y}=1$ and $t'=2$ in the model of $H_{3D}$, with darker and larger circles indicating larger normalized amplitudes, and color indicating sublattice localization. 
(a) SS0 hinge modes from stacks of S0 skin edge modes, with $\delta_{1,2,4}=-\delta_{3}=0.8$, and $\hat z$ couplings given by $t_\alpha=1$, $\delta_{a,c}=-\delta_{b,d}=0.8$; and (b) SSS corner modes from stacks of SS skin corner modes, with $\delta_{1,4}=-\delta_{2,3}=0.8$, $\hat z$ couplings given by $t_\alpha=1$, $\delta_{a,b,c,d}=0.8$.
}
\label{fig:3Dskin}
\end{figure}

To realize a higher-order SS0 skin-skin hinge mode, we stack 2D layers with skin edge modes (S0) from before [Fig.~2(e-g)], with symmetrized non-reciprocities $\delta_{1,2,4}=-\delta_3=0.8$. The 1D edge modes from each layer combine to form 2D ($\hat x$-$\hat z$) surface modes in the 3D system, and are pumped into 1D hinge modes by the surface skin effect induced by the nonzero net surface non-reciprocity. Indeed in Fig.~\ref{fig:3Dskin}(a), surface modes are driven toward the hinge at $(x,y)=(1,10)$, while those already on the hinge are also driven by the edge skin effect toward a corner, reminiscent of the cases in Fig.~4(b,c).
Similarity, the SSS skin modes can be realized by stacking 2D layers with skin corner modes (SS) of Fig.~2(b-d), as shown in Fig.~\ref{fig:3Dskin}(b). Note that in order to have skin effect along each direction, the net non-reciprocity along $z$ direction cannot destructively interfere in this case.

\end{document}